\documentclass[aps,prl,twocolumn,showpacs,amsmath,amssymbsuperscriptaddress]{revtex4-1}

\usepackage{amsmath,amsfonts,bm,graphicx,verbatim,mathrsfs,units}
\usepackage[colorlinks=true,citecolor=blue]{hyperref}

\newcommand{\aup}{\hat{a}^{\dag}}
\newcommand{\adown}{\hat{a}}
\newcommand{\cp}{\hat{c}^{\dag}}
\newcommand{\cdown}{\hat{c}}
\newcommand{\dup}{\hat{d}^{\dag}}
\newcommand{\ddown}{\hat{d}}

\begin{document}

\title{Hybrid Microwave-Cavity Heat Engine}

\author{Christian Bergenfeldt}
\affiliation{Physics Department, Lund University, Box 118, SE-22100 Lund, Sweden}
\author{Peter Samuelsson}
\affiliation{Physics Department, Lund University, Box 118, SE-22100 Lund, Sweden}
\author{Bj\"orn Sothmann}
\affiliation{D\'epartement de Physique Th\'eorique, Universit\'e de Gen\`{e}ve, CH-1211 Gen\`{e}ve 4, Switzerland}
\author{Christian Flindt}
\affiliation{D\'epartement de Physique Th\'eorique, Universit\'e de Gen\`{e}ve, CH-1211 Gen\`{e}ve 4, Switzerland}
\author{Markus B\"{u}ttiker}
\affiliation{D\'epartement de Physique Th\'eorique, Universit\'e de Gen\`{e}ve, CH-1211 Gen\`{e}ve 4, Switzerland}

\date{\today}

\begin{abstract}
We propose and analyze the use of hybrid microwave cavities as quantum heat engines. A possible realization consists of two macroscopically separated quantum dot conductors coupled capacitively to the fundamental mode of a microwave cavity. We demonstrate that an electrical current can be induced in one conductor through cavity-mediated processes by heating up the other conductor. The heat engine can reach Carnot efficiency with optimal conversion of heat to work. When the system delivers the maximum power, the efficiency can be a large fraction of the Carnot efficiency. The heat engine functions even with moderate electronic relaxation and dephasing in the quantum dots. We provide detailed estimates for the electrical current and output power using realistic parameters.
\end{abstract}


\maketitle

\emph{Introduction.---} Hybrid quantum systems that couple electronic transport in nano-scale conductors to photons in microwave cavities are currently going through a remarkable and rapid development. Several recent experiments have demonstrated controllable coupling of quantum dots to the fundamental mode of a microwave cavity \cite{Frey2011,Delbecq2011,Frey2012a,Frey2012b,Petersson2012,Delbecq2013,Toida2013,Wallraff2013,Basset2013}. These experimental advances are now paving the way for a broad spectrum of applications, ranging from hybrid quantum information processing \cite{Trif2008} and Cooper pair splitters \cite{Cottet2012} to on-chip micro-masers \cite{Childress2004} and quantum dot lasers \cite{Jin2011}. The ability to indirectly couple mesoscopic conductors over macroscopic distances via a microwave cavity \cite{Delbecq2013}, serving as a robust quantum bus for quantized energy and information flow, opens several intriguing avenues for the use of non-local electronic correlations mediated by cavity photons \cite{Bergenfeldt2012,Bergenfeldt2013,Lambert2013,Pulido2013,Xu2013,Xu2013b}.

\begin{figure}[h!]
\begin{center}
\includegraphics[width=0.9\columnwidth]{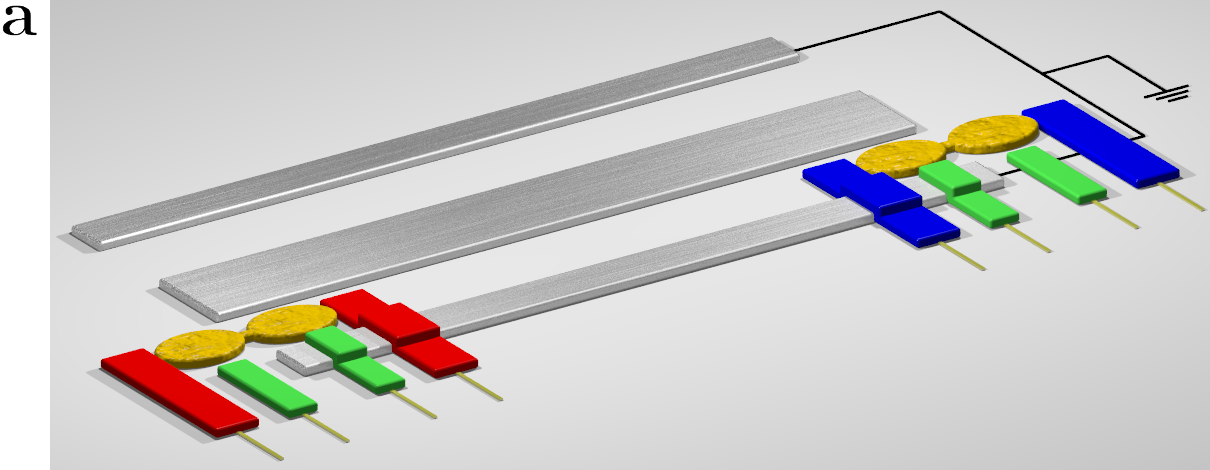}\vspace{0.5 cm}
\includegraphics[width=0.9\columnwidth]{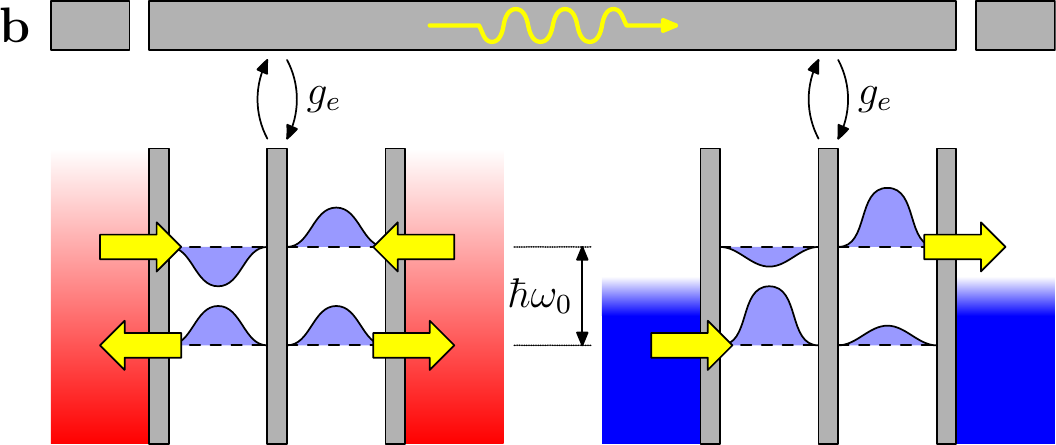}
\end{center}
\caption{(color online). Hybrid quantum heat engine. {\bf a}, Double quantum dots (yellow) coupled to each end of a microwave cavity (grey). External gates (green) are used to tune the quantum dots. The quantum dots are tunnel coupled to external electrodes.  Heat flows from the hot electrodes (red) to the cold electrodes (blue) via the double quantum dots and the microwave cavity. {\bf b}, The hybridized levels of the quantum dots are in resonance with the cavity frequency~$\omega_0$. Heat is transferred through processes, where electrons enter the excited state of the hot DQD and leave it via the ground state (arrows). With a finite mixing angle for the cold DQD, see Eq.~(\ref{eq:hyblevels}), electrons mainly enter the ground state from one lead and leave it from the excited state via the other lead. An electrical current is thereby induced in the cold conductor.
  \label{Fig1}}
\end{figure}

Parallel to these developments, research on nano-scale heat engines has witnessed several important advances. Here a central task is to direct energy from thermal fluctuations to electronic devices that are uncoupled from conventional power sources. To this end, quantum dots have emerged as promising candidates for nano-scale thermoelectrics  \cite{Beenakker1992,Humphrey2002,Esposito2009,Nakpathomkun2010,Entin2010,Sanchez2011,Sothmann2012,Sothmann2012b,Ruokola2012,Jordan2013,Kennes2013}. Experimentally, thermoelectric effects have been observed in two-terminal structures \cite{Staring1993,Dzurak1997,Scheibner2005,Scheibner2007,Svensson2012}. Further improvements are expected from three-terminal conductors that separate the input heat from the electrical output current \cite{Entin2010,Sanchez2011,Sothmann2012,Sothmann2012b,Ruokola2012,Jordan2013}. Still, these systems rely on a close proximity between the hot and the cold reservoirs which may lead to unwanted heat exchange that produces no work, but merely heats up the cold conductor \cite{Dresselhaus2007}. It is therefore desirable to separate the hot and the cold conductors and have the heat current flow between them in a highly controllable way.

In this Letter we propose and analyze the use of hybrid cavity-QED systems as quantum heat engines. Our idea can be implemented in a variety of system architectures, but to be specific we consider the setup depicted in Fig.~\ref{Fig1}a: Two double quantum dots (DQDs) coupled to external electrodes interact capacitively with each end of a microwave cavity of macroscopic dimensions (around 1~cm in recent experiments \cite{Frey2011,Delbecq2011,Frey2012a,Frey2012b,Petersson2012,Delbecq2013,Toida2013,Wallraff2013,Basset2013}) which serves as a quantum bus for heat currents between them \cite{Meschke2006}. Due to the large distance between the conductors other types of heat exchange, e.~g.~due to phonons, are negligible \cite{Savin2006}. We heat up one of the conductors and establish an energy flow from the hot conductor to the cold conductor via the microwave cavity. An asymmetry in the cold conductor makes it possible to rectify the thermal fluctuations due the heat current such that a directed electrical current is induced. Rectification is achieved by having the cold conductor couple more strongly to one of the electrodes at higher energies and more strongly to the other at lower energies as detailed below. As we go on to show, the heat engine can reach Carnot efficiency where the conversion from heat to work is optimal. Moreover, at maximum power a sizable fraction of the Carnot efficiency is achievable.  Importantly, the heat engine may function efficiently even under the influence of electronic dephasing and relaxation as we will see.

\emph{Heat engine.---} Our system is shown schematically in Fig.~\ref{Fig1}b. Due to strong Coulomb interactions the DQDs are either empty or occupied by a single electron. The spin degree of freedom is ignored as it would only renormalize the tunneling rates found below. The hybrized levels of the DQDs $(i=1,2)$ can be written in terms of the left ($|L\rangle_i$) and right ($|R\rangle_i$) DQD states as
\begin{equation}
\label{eq:hyblevels}
\begin{split}
|+\rangle_i&=\cos(\theta_{i})|L\rangle_i-\sin(\theta_{i})|R\rangle_i,\\
|-\rangle_i&=\sin(\theta_{i})|L\rangle_i+\cos(\theta_{i})|R\rangle_i,
\end{split}
\end{equation}
where $\theta_{i}=\arctan[2 t_i/(\sqrt{\varepsilon_i^2+(2t_i)^2}+\varepsilon_i)]$ are the mixing angles, given by the tunnel couplings $t_i$ and the energy dealignments $\varepsilon_i$ of the localized levels. The mixing angles can be electrostatically controlled using external gates and they can be chosen such that the ground state ($|-\rangle_i$) couples more strongly to one electrode and the excited state ($|+\rangle_i$) more strongly to the other, see Fig.~\ref{Fig1}b.

Independently of the mixing angles, the energy splitting of the hybridized levels can be tuned into resonance with the fundamental mode of the microwave cavity with frequency $\omega_0$, such that  $\hbar\omega_0=\sqrt{\varepsilon_i^2+(2t_i)^2}$.  For regular superconducting transmission line cavities (typically made of Al or Nb) the characteristic impedance $Z_{0}$ is much smaller than the resistance quantum $R_{Q}=h/e^{2}$. The DQDs-cavity system itself can then be described by
the generalized Tavis-Cummings Hamiltonian~\cite{Bergenfeldt2013}
\begin{equation}
\hat{H}_{S}=\hbar\omega_{0}\aup\adown+\sum_{i=1,2}\left[\hbar\omega_{0}\hat{\Delta}_i+\hbar g_{e}(\aup\dup_{-i}\ddown_{+i}+\mathrm{h.c.})\right],
\label{eq:Tavis}
\nonumber
\end{equation}
where $\aup$ creates excitations of the cavity mode. The creation operator for the ground (excited) state of the DQDs is written as $\dup_{-(+)i}$ and $\hat{\Delta}_i= (\hat{n}_{+i}-\hat{n}_{-i})/2$, where $\hat{n}_{\pm i}=\dup_{\pm i}\ddown_{\pm i}$ are number operators. The effective couplings between the hybridized levels and the cavity are given in terms of the bare coupling $g_i$ as $g_e=(t_i/\hbar\omega_0) g_i$ and are for simplicity taken equal for the two DQDs.

Each quantum dot is tunnel coupled to an electronic lead in local thermal equilibrium. The leads connected to the same DQD are kept at the temperature $T_{i}$ and $\mu_{\nu i}$ is the chemical potential of the left ($\nu=L$) or right ($\nu=R$) lead. We describe the leads by the Hamiltonian $\hat{H}_{\nu}=\sum_{k \nu i}\epsilon_{k}\cp_{k\nu i}\cdown_{k\nu i}$, where $\cp_{k\nu i}$ creates an electron at energy $\epsilon_{k}$ in lead $\nu$ connected to DQD $i$. Finally, the coupling between the leads and the DQDs is governed by the tunneling Hamiltonian
\begin{equation}
\hat{H}_{T}\!=\!\! \sum_{k i}\left[
                            \begin{array}{c}
                              t_{kLi}\cdown_{kLi} \\
                              t_{kRi}\cdown_{kRi} \\
                            \end{array}
                          \right]^T
                          \left[
                            \begin{array}{cc}
                              \cos(\theta_i) & \sin(\theta_i)\\
                              -\sin(\theta_i) & \cos(\theta_i)\\
                            \end{array}
                          \right]
                          \left[
                            \begin{array}{c}
                              \dup_{+i} \\
                              \dup_{-i} \\
                            \end{array}
                          \right]\!\!+\mathrm{h.c.}.
\nonumber
\end{equation}

\begin{figure*}
\includegraphics[width=0.83\textwidth]{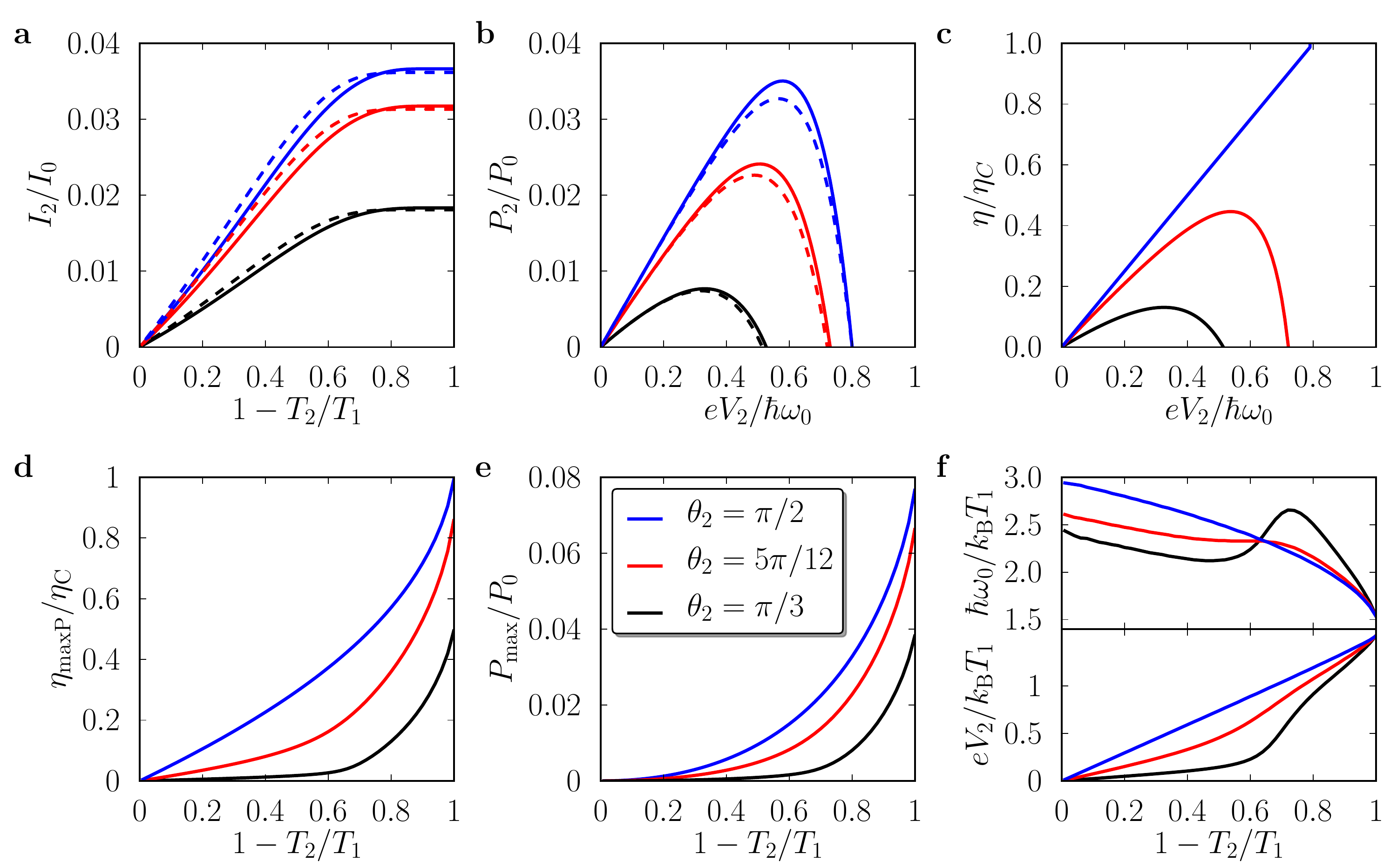}
	\caption{\label{fig:current} (color online). Thermoelectric performance. {\bf a}, Thermally induced current in DQD 2 as a function of the temperature ratio $T_2/T_1$ with different mixing angles $\theta_{2}$. Here $k_{B}T_{1}=0.5\hbar\omega_{0}$ and $I_{0}=e\Gamma/2$ with $\Gamma=\Gamma_1=\Gamma_2$. Full numerics (continuous lines) compare well with Eq.~(\ref{eq:current_analytic}) (dashed lines). {\bf b}, Power as a function of the applied voltage $V_2$. Temperatures are $k_{B}T_{1}=0.5\hbar\omega_{0}$ and  $k_{B}T_{2}=0.1\hbar\omega_{0}$, and $P_{0}=\Gamma k_{B}T_{1}$. Equation (\ref{powerana}) is shown with a dashed line. {\bf c}, Efficiency $\eta$ over the Carnot efficiency $\eta_C$ versus the applied voltage $V_2$. Same parameters as in {\bf b}. {\bf d}, Efficiency at maximum power as a function of the temperature ratio $T_2/T_1$. {\bf e}, Corresponding maximum power. {\bf f}, Optimized values of the voltage $V_2$ and the frequency $\omega_0$.}
\end{figure*}

\emph{Quantum master equation.---} Electron transport in the DQDs occurs via sequential tunneling events described by a quantum master equation (QME) for the reduced density operator $\hat{\rho}$ of the DQDs and the cavity. By integrating out the leads in the standard Born-Markov approximation we arrive at the QME~\cite{Bergenfeldt2013,Breuer2002}
\begin{equation}
\frac{d}{dt}\hat{\rho}=\mathcal{L}[\hat{\rho}]=-\frac{i}{\hbar}[\hat{H}_{S},\hat{\rho}]+\mathcal{L}_{\rm elec}[\hat{\rho}].
\label{QME}
\end{equation}
The commutator corresponds to the coherent evolution due to $\hat{H}_{S}$ and $\mathcal{L}_{\rm elec}=\sum_{i,\nu,\xi=\pm}\mathcal{L}_{i\nu\xi}$ describes tunneling events between the electronic leads and the DQDs. The tunneling rates $\Gamma_{\nu i}=2\pi\sum_{k}|t_{k\nu i}|^{2}\delta(\epsilon-\epsilon_{k})$, $\nu=L,R$, $i=1,2$, are energy-independent and chosen symmetrically for each DQD, $\Gamma_{Li}=\Gamma_{Ri}=\Gamma_{i}$. We consider the experimentally relevant regime $\{\omega_0; k_BT_i/\hbar\}\gg \{g_e; \Gamma_{ i}\}\gg\kappa$, where the cavity decay rate $\kappa$ is typically so small that it can be neglected below \cite{Wang2009}. The number of photons in the cavity is then determined by the electronic transport in the DQDs. Electron tunneling is accounted for by the Lindblad terms $\mathcal{L}_{i\nu\xi}[\hat{\rho}]=(\bar{\Gamma}_{i\nu\xi }(\theta_{i})/2)\{f_{i\nu}(\epsilon_{\xi})\mathcal{D}[\hat d_{i \xi },\hat{\rho}]+[1-f_{i
\nu }(\epsilon_{\xi})]\mathcal{D}[\hat d^{\dagger}_{i \xi},\hat{\rho}]\}$ with the dissipator $\mathcal{D}[\hat{\gamma},\hat{\rho}]=2\hat{\gamma}^{\dag}\hat{\rho}\hat{\gamma}-\{\hat{\gamma}\hat{\gamma}^{\dag},\hat{\rho}\}$ \cite{Bergenfeldt2013}. We have also defined $\epsilon_{\pm}=\pm\hbar\omega_0/2$, $\bar{\Gamma}_{iL-(R+)}(\theta_{i})=\Gamma_{i}\sin^{2}(\theta_{i})$,  and $\bar{\Gamma}_{iL+(R-)}(\theta_{i})=\Gamma_{i}\cos^{2}(\theta_{i})$, and $f_{\nu i}(\epsilon)$ is the Fermi distribution of lead $\nu=L,R$ coupled to DQD $i=1,2$. The dependence of the tunneling rates $\bar{\Gamma}_{i\nu\xi}$ on the mixing angles reflects the asymmetry of the hybrized DQD states.

We first analyze the ideal situation without electronic dephasing and relaxation in the DQDs, before discussing these important issues in detail. We also start out in the strong coupling limit, $g_{e}\gg\Gamma_{i}$, but later relax this assumption. For strong couplings, a secular approximation allows us to neglect coherences between non-degenerate states of $\hat{H}_{S}$ and the QME reduces to an ordinary master equation \cite{Bergenfeldt2013}, which eases the analysis below.

\emph{Thermoelectrics.---} To evaluate the performance of the heat engine we identify the (super) operators for the charge and heat currents. The operator $\mathcal{I}_{i}$ for the charge current through the right lead of DQD $i$ acting on a density matrix $\hat{\rho}$ reads $\mathcal{I}_{i}[\hat{\rho}]=e\sum_{\xi}\bar
\Gamma_{iR\xi}(\theta_i)([1-f_{Ri}(\epsilon_{\xi})]\hat d^{\dagger}_{\xi i}\hat{\rho}\hat d_{\xi i}-f_{Ri}(\epsilon_{\xi})\hat d_{\xi i}\hat{\rho}\hat d^{\dagger}_{\xi i})$. In the stationary state, defined by $\mathcal{L}[\hat{\rho}^{\rm stat}]=0$ and $\mathrm{Tr}\{\hat{\rho}^{\rm stat}\}$=1, the average charge current is
$I_{i}=\mathrm{Tr}\{\mathcal{I}_{i}[\hat{\rho}^{\rm stat}]\}$. The operator $\mathcal{J}_{\nu i}$ for the heat current in lead $\nu=L,R$ connected to DQD $i$ reads $\mathcal{J}_{\nu i}[\hat{\rho}]=\sum_{\xi}(\epsilon_{\xi}-\mu_{\nu i})\bar \Gamma_{i\nu\xi}(\theta_i)([1-f_{\nu i}(\epsilon_{\xi})]\hat d^{\dagger}_{\xi i}\hat{\rho}\hat d_{\xi i}-f_{\nu i}(\epsilon_{\xi})\hat d_{\xi i}\hat{\rho}\hat d^{\dagger}_{\xi i})$ and the average heat current is $J_{\nu i}=\mathrm{Tr}\{\mathcal{J}_{\nu i}[\hat{\rho}^{\rm stat}]\}$.

To begin with we apply no voltages. Instead, we analyze how an electrical current is induced by heating up the leads connected to DQD 1. A finite level detuning in one of the DQDs is required to induce a current. We take $\varepsilon_1=0$ ($\theta_{1}=\pi/4$), such that no current is induced in DQD 1. On the other hand, by having a finite detuning in DQD 2 ($\varepsilon_2\neq0$ and hence $\theta_{2}\neq \pi/4$), the excited state of DQD 2 couples more strongly to one electrode and the ground state more strongly to the other. Electrons then preferably tunnel into the ground state of DQD 2 from one electrode, absorb a photon from the cavity, bringing it to the excited state, and finally leave the DQD via the other electrode, Fig.~\ref{Fig1}b. On average, photons are emitted from DQD 1 and absorbed by DQD 2, where a net charge current is generated. Unlike proposals relying on energy-dependent tunneling barriers \cite{Sanchez2011,Sothmann2012}, the asymmetry due to a finite level detuning can be externally controlled. The direction of the current is determined by having $\theta_{2}<\pi/4$ or $\theta_{2}>\pi/4$. An electrical current can also be induced by heating DQD 2 (or cooling DQD 1), but we will not consider this option further. The heat flow between the DQDs may be controlled by bringing them in and out of resonance with the cavity \cite{Lambert2013,Pulido2013}.

Figure~\ref{fig:current} summarizes our thermoelectric analysis of the heat engine. Figure 2{\bf a} shows  the thermally induced charge current $I_2$ in DQD 2 as a function of the ratio of temperatures $T_2/T_1$. The current vanishes at equilibrium, $T_2=T_1$, but increases as DQD 1 is heated up and eventually saturates at large temperature differences, $T_1\gg T_2$. As $\theta_2\rightarrow\pi/2$, the thermoelectric tight-coupling limit is approached, where each photon absorbed from the cavity leads to the transfer of exactly one electron through DQD 2. The ratio of the electrical current in DQD 2 over the input heat current $J_{1}=J_{L1}+J_{R1}$ from DQD 1 is then simply $I_2/J_1=e/\hbar\omega_0$, which is the ratio of the electronic charge over the energy quantum \footnote{The case $\theta_2\rightarrow\pi/2$ should be treated with care as the effective couplings to the cavity vanish for $\theta_2=\pi/2$ and we go outside the strong coupling regime assumed so far.}.  For $\bar{f}_{i}=f_{L i}(\hbar\omega_0/2)=f_{R i}(\hbar\omega_0/2)\ll 1$, where the cavity is mostly empty, the electrical current is well-approximated by the analytic expression
\begin{equation}
I_{2}=\cos(2\theta_{2})\frac{e\Gamma_{1}\Gamma_{2}}{\Gamma_{1}+\Gamma_{2}}(\bar{f}_{2}^{2}-\bar{f}_{1}^{2}).
\label{eq:current_analytic}
\end{equation}
The electrical current is maximal for equal tunneling rates and we therefore proceed with $\Gamma=\Gamma_1=\Gamma_2$.

\emph{Power \& efficiency.---} To extract power from the heat engine, a bias voltage $V_2$ must be applied against the heat-induced charge current. Figure 2{\bf b} shows the power $P_2=I_2V_2$. The power vanishes at $V_2=0$ as well as at the stopping voltage $V_\text{stop}$, where the heat-induced and the bias-driven currents compensate each other, and reaches a maximum in between. Due to the nonlinear current-temperature characteristics in Fig.~2{\bf a}, the maximum power does not occur at $V_2=V_\text{stop}/2$, but is shifted to larger values. For $\bar{f}_{\nu i}=f_{\nu  i}(\hbar\omega_0/2)\ll1$ with $\mu_{(R/L)2}=\pm eV_2/2$, the analytical expression
\begin{equation}
\begin{split}
P_2=\frac{e\Gamma V_2}{2}\Bigg[&\sin(2\theta_{2})\{\bar{f}_{L2}-\bar{f}_{R2}\}-\cos(2\theta_{2})\bar{f}_{1}^{2}\\
&+\cos(2\theta_{2})\bar{f}_{L2}\bar{f}_{R2}+\frac{\sin^{2}(2\theta_{2})}{4}\{\bar{f}_{L2}^{2}-\bar{f}_{R2}^{2}\}\Bigg]
\end{split}
\label{powerana}
\end{equation}
is useful for further performance optimization below.

The efficiency of the heat engine is quantified by the ratio of the output power over the input heat, $\eta=P_2/J_1$, Fig.~2{\bf c}. In general, the efficiency grows upon increasing the voltage and it reaches a maximum before dropping to $\eta=0$ at the stopping voltage. Remarkably, the largest efficiency occurs roughly at the same voltage as the maximum power. In the tight-coupling limit, $\theta_2=\pi/2$, the efficiency shows a qualitatively different behavior: It grows linearly with the voltage and reaches Carnot efficiency $\eta_C=1-T_2/T_1$ at the stopping voltage. At this point, however, the heat engine operates reversibly and produces no output power. Figure~2{\bf d}, instead, shows the efficiency at \emph{maximum} power $\eta_\text{maxP}$. We note that the results in the tight-coupling limit provide a theoretical upper limit on the efficiency and the efficiency at maximum power in Figs.~2{\bf c} and 2{\bf d}. The maximum power as well as the optimized values of $V_2$ and $\omega_0$ are shown in Figs.~2{\bf e} and 2{\bf f}, respectively. In the tight-coupling limit, $\eta_\text{maxP}$ grows as $\eta_C/2$ for small temperature differences in agreement with general thermodynamic bounds for systems with time-reversal symmetry~\cite{Broeck2005}. The efficiency at maximum power grows strongly with increasing temperature differences and it  reaches $\eta_C$ with $T_1\gg T_2$. It satisfies the bounds $\eta_C/2\leq\eta_\text{maxP}\leq \eta_C/(2-\eta_C)$~\cite{Schmiedl2008}. For $\theta_2<\pi/2$, the efficiency at maximum power is slightly reduced, but shows similar behavior. Importantly, the heat engine has thermoelectric properties that compare well with other systems \cite{Sanchez2011,Sothmann2012}, but here the hot and cold electrodes are separated by macroscopic distances.

\begin{figure}
\includegraphics[width=0.9\columnwidth]{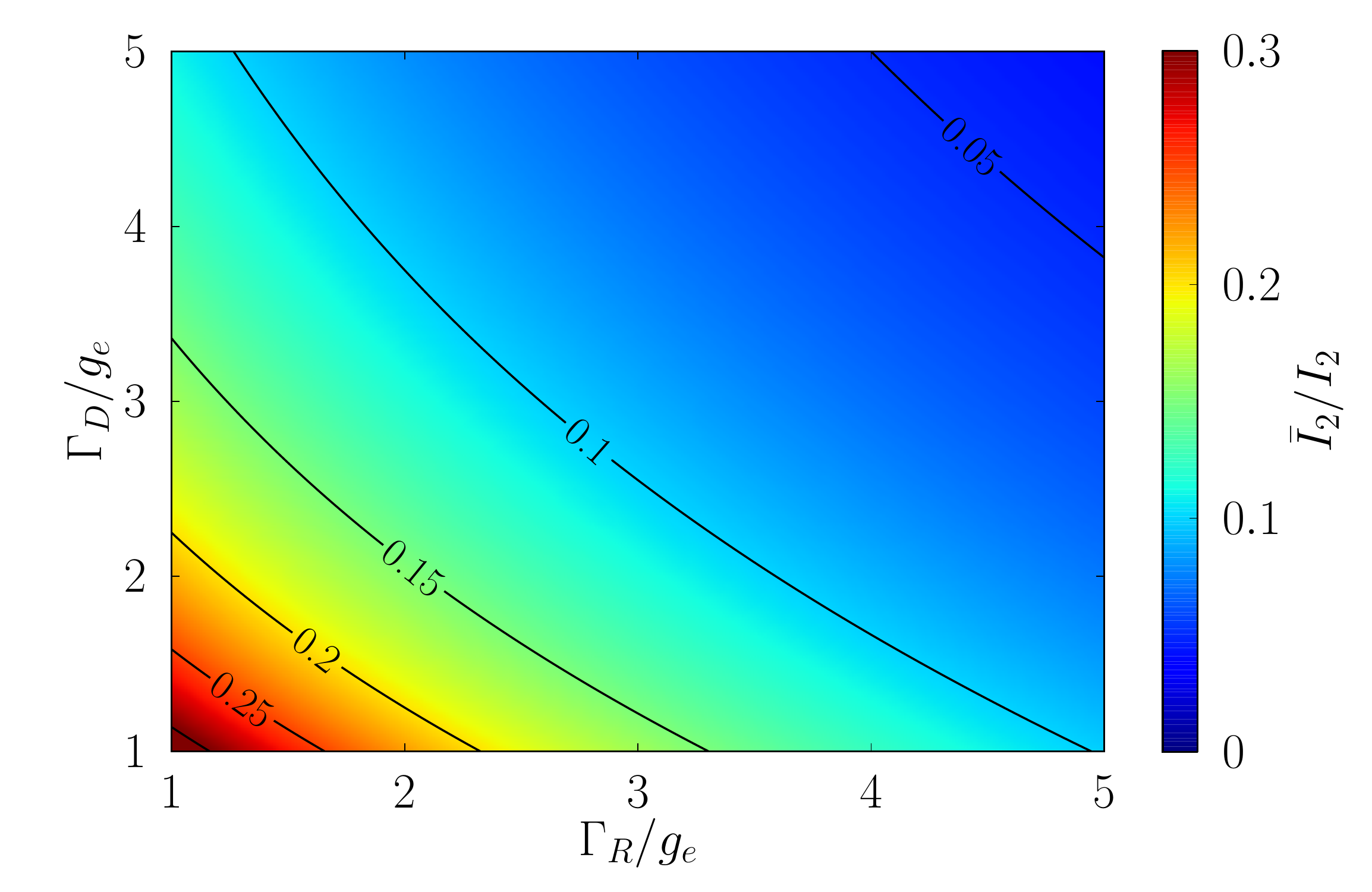}
	\caption{\label{depplot} (color online). Influence of electronic dephasing and relaxation. Thermally induced current in DQD 2 as a function of the electronic relaxation and dephasing rates $\Gamma_R$ and $\Gamma_D$. Under optimal conditions, the current takes on the value $I_{0}=e\Gamma/2$. \label{Fig3}}
\end{figure}

\emph{Relaxation \& dephasing.---} We now turn to the influence of electronic dephasing and relaxation in the DQDs. We also relax the assumption of strong couplings, $g_{e}\gg \Gamma_{i} $. Beyond the strong-coupling limit, the interaction time between electrons on the DQDs and the cavity photons is reduced. Electronic relaxation and dephasing in the DQDs are accounted for by adding the terms \cite{Breuer2002}
\begin{equation}
\mathcal{L}_{R}[\hat{\rho}]=\frac{\Gamma_{R}}{2}\sum_{i}\left(\{1-\bar{f}_{i}^{2}\}\mathcal{D}[\dup_{+i}\ddown_{-i},\hat{\rho}]+\bar{f}_{i}^{2}\mathcal{D}[\dup_{-i}\ddown_{+i},\hat{\rho}]\right)
\nonumber
\end{equation}
and
\begin{equation}
\mathcal{L}_{D}[\hat{\rho}]=\frac{\Gamma_{D}}{2}\sum_{i}\mathcal{D}[\dup_{+i}\ddown_{+i}-\dup_{-i}\ddown_{-i},\hat{\rho}]
\nonumber
\end{equation}
to the QME~\eqref{QME}. For finite relaxation and dephasing rates, $\Gamma_{R}$ and $\Gamma_{D}$, together with $\Gamma_{i}\sim g_{e}$, the mean current in Eq.~\eqref{eq:current_analytic} is modified as
\begin{equation}
\bar{I}_{2}=\frac{4 g_{e}^{2}}{(\Gamma+\Gamma_{R})(\Gamma+\Gamma_{R}+4\Gamma_{D})+4g_{e}^{2}}I_{2}.
\label{eq:reCurr}
\end{equation}
With $\Gamma_{D}=\Gamma_{R}=0$, the current and the power are maximal for $\Gamma=2g_{e}$, which gives $\bar{I}_{2}=I_{2}/2$. Figure~\ref{Fig3} shows that a considerable current is achievable even with moderate relaxation and dephasing rates.

\emph{Estimates.---} Finally, we can provide estimates for the current and power produced by our heat engine. We consider a standard transmission line cavity with frequency $\omega_{0}\simeq 2\pi\times\unit[10]{GHz}$ and impedance $Z_{0}=\unit[50]{\Omega}$.  The effective cavity coupling is $g_{e}\simeq \sin(2\theta_{2})\omega_{0}\sqrt{2Z_{0}/R_{Q}}$ \cite{Childress2004,Bergenfeldt2012,Bergenfeldt2013}. With $\theta_{2}=\pi/3$ we find $g_{e}\simeq\unit[0.5]{GHz}$. For the electronic leads we take $\Gamma=2g_{e}\simeq \unit[1]{GHz}$ together with reasonable temperatures $T_1= 2 T_{2}\simeq \hbar\omega_0/k_{B}\sim \unit[0.2]{K}$. From Fig.~\ref{fig:current} we then obtain the conservative estimates  $\bar{I}_{2}\simeq\unit[1]{pA}$ and $P_{\mathrm{max}}\simeq\unit[0.1]{fW}$. These figures compare well with existing proposals for heat engines operating in the Coulomb blockage regime, see e.~g.~Ref.~\cite{Sothmann2012}.  Further optimization may enhance the values. Realistic relaxation and dephasing rates, $\Gamma_{R}\simeq\Gamma_{D}\simeq\unit[1]{GHz}$~\cite{Hayashi2003}, give $\Gamma_{D}/g_{e}\simeq \Gamma_{R}/g_{e} \simeq 2$, which would roughly reduce the mean current by a factor of 5 according to Fig.~\ref{Fig3}.

\emph{Conclusions.---}
We have proposed and analyzed the use of hybrid microwave cavities as quantum heat engines. A possible realization of our ideas consists of DQDs capacitively coupled to a microwave cavity. By heating up one of the DQDs, a heat current through the cavity can induce an electrical current in the other DQD. At maximum power, the efficiency of the heat engine can be a large fraction of the Carnot efficiency. Moreover, the heat engine can operate even with moderate electronic relaxation and dephasing in the quantum dots.

\emph{Acknowledgements.---} We thank R.~S\'{a}nchez for useful comments on the manuscript. The work is supported by ESF, Swedish Research Council, NanoPower, Swiss NSF, and the NCCR QSIT.


\begin{thebibliography}{40}%
\makeatletter
\providecommand \@ifxundefined [1]{%
 \@ifx{#1\undefined}
}%
\providecommand \@ifnum [1]{%
 \ifnum #1\expandafter \@firstoftwo
 \else \expandafter \@secondoftwo
 \fi
}%
\providecommand \@ifx [1]{%
 \ifx #1\expandafter \@firstoftwo
 \else \expandafter \@secondoftwo
 \fi
}%
\providecommand \natexlab [1]{#1}%
\providecommand \enquote  [1]{``#1''}%
\providecommand \bibnamefont  [1]{#1}%
\providecommand \bibfnamefont [1]{#1}%
\providecommand \citenamefont [1]{#1}%
\providecommand \href@noop [0]{\@secondoftwo}%
\providecommand \href [0]{\begingroup \@sanitize@url \@href}%
\providecommand \@href[1]{\@@startlink{#1}\@@href}%
\providecommand \@@href[1]{\endgroup#1\@@endlink}%
\providecommand \@sanitize@url [0]{\catcode `\\12\catcode `\$12\catcode
  `\&12\catcode `\#12\catcode `\^12\catcode `\_12\catcode `\%12\relax}%
\providecommand \@@startlink[1]{}%
\providecommand \@@endlink[0]{}%
\providecommand \url  [0]{\begingroup\@sanitize@url \@url }%
\providecommand \@url [1]{\endgroup\@href {#1}{\urlprefix }}%
\providecommand \urlprefix  [0]{URL }%
\providecommand \Eprint [0]{\href }%
\@ifxundefined \urlstyle {%
  \providecommand \doi  [0]{\begingroup \@sanitize@url \@doi}%
  \providecommand \@doi [1]{\endgroup \@@startlink {\doibase
  #1}doi:\discretionary {}{}{}#1\@@endlink }%
}{%
  \providecommand \doi  [0]{doi:\discretionary{}{}{}\begingroup
  \urlstyle{rm}\Url }%
}%
\providecommand \doibase [0]{http://dx.doi.org/}%
\providecommand \Doi [0]{\begingroup \@sanitize@url \@Doi }%
\providecommand \@Doi  [1]{\endgroup\@@startlink{\doibase#1}\@@Doi}%
\providecommand \@@Doi [1]{#1\@@endlink}%
\providecommand \selectlanguage [0]{\@gobble}%
\providecommand \bibinfo  [0]{\@secondoftwo}%
\providecommand \bibfield  [0]{\@secondoftwo}%
\providecommand \translation [1]{[#1]}%
\providecommand \BibitemOpen [0]{}%
\providecommand \bibitemStop [0]{}%
\providecommand \bibitemNoStop [0]{.\EOS\space}%
\providecommand \EOS [0]{\spacefactor3000\relax}%
\providecommand \BibitemShut  [1]{\csname bibitem#1\endcsname}%
\bibitem [{\citenamefont {Frey}\ \emph {et~al.}(2011)\citenamefont {Frey},
  \citenamefont {Leek}, \citenamefont {Beck}, \citenamefont {Ensslin},
  \citenamefont {Wallraff},\ and\ \citenamefont {Ihn}}]{Frey2011}%
  \BibitemOpen
  \bibfield  {author} {\bibinfo {author} {\bibfnamefont {T.}~\bibnamefont
  {Frey}}, \bibinfo {author} {\bibfnamefont {P.~J.}\ \bibnamefont {Leek}},
  \bibinfo {author} {\bibfnamefont {M.}~\bibnamefont {Beck}}, \bibinfo {author}
  {\bibfnamefont {K.}~\bibnamefont {Ensslin}}, \bibinfo {author} {\bibfnamefont
  {A.}~\bibnamefont {Wallraff}}, \ and\ \bibinfo {author} {\bibfnamefont
  {T.}~\bibnamefont {Ihn}},\ }\Doi {10.1063/1.3604784} {\bibfield  {journal}
  {\bibinfo  {journal} {Appl. Phys. Lett.} }\textbf {\bibinfo {volume}
  {98}},\ \bibinfo {pages} {262105} (\bibinfo {year} {2011})}\BibitemShut
  {NoStop}%
\bibitem [{\citenamefont {Delbecq}\ \emph {et~al.}(2011)\citenamefont
  {Delbecq}, \citenamefont {Schmitt}, \citenamefont {Parmentier}, \citenamefont
  {Roch}, \citenamefont {Viennot}, \citenamefont {F\`eve}, \citenamefont
  {Huard}, \citenamefont {Mora}, \citenamefont {Cottet},\ and\ \citenamefont
  {Kontos}}]{Delbecq2011}%
  \BibitemOpen
  \bibfield  {author} {\bibinfo {author} {\bibfnamefont {M.~R.}\ \bibnamefont
  {Delbecq}}, \bibinfo {author} {\bibfnamefont {V.}~\bibnamefont {Schmitt}},
  \bibinfo {author} {\bibfnamefont {F.~D.}\ \bibnamefont {Parmentier}},
  \bibinfo {author} {\bibfnamefont {N.}~\bibnamefont {Roch}}, \bibinfo {author}
  {\bibfnamefont {J.~J.}\ \bibnamefont {Viennot}}, \bibinfo {author}
  {\bibfnamefont {G.}~\bibnamefont {F\`eve}}, \bibinfo {author} {\bibfnamefont
  {B.}~\bibnamefont {Huard}}, \bibinfo {author} {\bibfnamefont
  {C.}~\bibnamefont {Mora}}, \bibinfo {author} {\bibfnamefont {A.}~\bibnamefont
  {Cottet}}, \ and\ \bibinfo {author} {\bibfnamefont {T.}~\bibnamefont
  {Kontos}},\ }\Doi {10.1103/PhysRevLett.107.256804} {\bibfield  {journal}
  {\bibinfo  {journal} {Phys. Rev. Lett.} }\textbf {\bibinfo {volume}
  {107}},\ \bibinfo {pages} {256804} (\bibinfo {year} {2011})}\BibitemShut
  {NoStop}%
\bibitem [{\citenamefont {Frey}\ \emph
  {et~al.}(2012){\natexlab{a}}\citenamefont {Frey}, \citenamefont {Leek},
  \citenamefont {Beck}, \citenamefont {Blais}, \citenamefont {Ihn},
  \citenamefont {Ensslin},\ and\ \citenamefont {Wallraff}}]{Frey2012a}%
  \BibitemOpen
  \bibfield  {author} {\bibinfo {author} {\bibfnamefont {T.}~\bibnamefont
  {Frey}}, \bibinfo {author} {\bibfnamefont {P.~J.}\ \bibnamefont {Leek}},
  \bibinfo {author} {\bibfnamefont {M.}~\bibnamefont {Beck}}, \bibinfo {author}
  {\bibfnamefont {A.}~\bibnamefont {Blais}}, \bibinfo {author} {\bibfnamefont
  {T.}~\bibnamefont {Ihn}}, \bibinfo {author} {\bibfnamefont {K.}~\bibnamefont
  {Ensslin}}, \ and\ \bibinfo {author} {\bibfnamefont {A.}~\bibnamefont
  {Wallraff}},\ }\Doi {10.1103/PhysRevLett.108.046807} {\bibfield  {journal}
  {\bibinfo  {journal} {Phys. Rev. Lett.} }\textbf {\bibinfo {volume}
  {108}},\ \bibinfo {pages} {046807} (\bibinfo {year}
  {2012}{\natexlab{a}})}\BibitemShut {NoStop}%
\bibitem [{\citenamefont {Frey}\ \emph
  {et~al.}(2012){\natexlab{b}}\citenamefont {Frey}, \citenamefont {Leek},
  \citenamefont {Beck}, \citenamefont {Faist}, \citenamefont {Wallraff},
  \citenamefont {Ensslin}, \citenamefont {Ihn},\ and\ \citenamefont
  {B\"uttiker}}]{Frey2012b}%
  \BibitemOpen
  \bibfield  {author} {\bibinfo {author} {\bibfnamefont {T.}~\bibnamefont
  {Frey}}, \bibinfo {author} {\bibfnamefont {P.~J.}\ \bibnamefont {Leek}},
  \bibinfo {author} {\bibfnamefont {M.}~\bibnamefont {Beck}}, \bibinfo {author}
  {\bibfnamefont {J.}~\bibnamefont {Faist}}, \bibinfo {author} {\bibfnamefont
  {A.}~\bibnamefont {Wallraff}}, \bibinfo {author} {\bibfnamefont
  {K.}~\bibnamefont {Ensslin}}, \bibinfo {author} {\bibfnamefont
  {T.}~\bibnamefont {Ihn}}, \ and\ \bibinfo {author} {\bibfnamefont
  {M.}~\bibnamefont {B\"uttiker}},\ }\Doi {10.1103/PhysRevB.86.115303}
  {\bibfield  {journal} {\bibinfo  {journal} {Phys. Rev. B} }\textbf
  {\bibinfo {volume} {86}},\ \bibinfo {pages} {115303} (\bibinfo {year}
  {2012}{\natexlab{b}})}\BibitemShut {NoStop}%
\bibitem [{\citenamefont {Petersson}\ \emph {et~al.}(2012)\citenamefont
  {Petersson}, \citenamefont {McFaul}, \citenamefont {Schroer}, \citenamefont
  {Jung}, \citenamefont {Taylor}, \citenamefont {Houck},\ and\ \citenamefont
  {Petta}}]{Petersson2012}%
  \BibitemOpen
  \bibfield  {author} {\bibinfo {author} {\bibfnamefont {K.~D.}\ \bibnamefont
  {Petersson}}, \bibinfo {author} {\bibfnamefont {L.~W.}\ \bibnamefont
  {McFaul}}, \bibinfo {author} {\bibfnamefont {M.~D.}\ \bibnamefont {Schroer}},
  \bibinfo {author} {\bibfnamefont {M.}~\bibnamefont {Jung}}, \bibinfo {author}
  {\bibfnamefont {J.~M.}\ \bibnamefont {Taylor}}, \bibinfo {author}
  {\bibfnamefont {A.~A.}\ \bibnamefont {Houck}}, \ and\ \bibinfo {author}
  {\bibfnamefont {J.~R.}\ \bibnamefont {Petta}},\ }\Doi {10.1038/nature11559}
  {\bibfield  {journal} {\bibinfo  {journal} {Nature} }\textbf {\bibinfo
  {volume} {490}},\ \bibinfo {pages} {380} (\bibinfo {year}
  {2012})}\BibitemShut {NoStop}%
\bibitem [{\citenamefont {Delbecq}\ \emph {et~al.}(2013)\citenamefont
  {Delbecq}, \citenamefont {Bruhat}, \citenamefont {Viennot}, \citenamefont
  {Datta}, \citenamefont {Cottet},\ and\ \citenamefont {Kontos}}]{Delbecq2013}%
  \BibitemOpen
  \bibfield  {author} {\bibinfo {author} {\bibfnamefont {M.~R.}\ \bibnamefont
  {Delbecq}}, \bibinfo {author} {\bibfnamefont {L.~E.}\ \bibnamefont {Bruhat}},
  \bibinfo {author} {\bibfnamefont {J.~J.}\ \bibnamefont {Viennot}}, \bibinfo
  {author} {\bibfnamefont {S.}~\bibnamefont {Datta}}, \bibinfo {author}
  {\bibfnamefont {A.}~\bibnamefont {Cottet}}, \ and\ \bibinfo {author}
  {\bibfnamefont {T.}~\bibnamefont {Kontos}},\ }\Doi {10.1038/ncomms2407}
  {\bibfield  {journal} {\bibinfo  {journal} {Nature Commun.} }\textbf
  {\bibinfo {volume} {4}},\ \bibinfo {pages} {1400} (\bibinfo {year}
  {2013})}\BibitemShut {NoStop}%
\bibitem [{\citenamefont {Toida}\ \emph {et~al.}(2013)\citenamefont {Toida},
  \citenamefont {Nakajima},\ and\ \citenamefont {Komiyama}}]{Toida2013}%
  \BibitemOpen
  \bibfield  {author} {\bibinfo {author} {\bibfnamefont {H.}~\bibnamefont
  {Toida}}, \bibinfo {author} {\bibfnamefont {T.}~\bibnamefont {Nakajima}}, \
  and\ \bibinfo {author} {\bibfnamefont {S.}~\bibnamefont {Komiyama}},\ }\Doi
  {10.1103/PhysRevLett.110.066802} {\bibfield  {journal} {\bibinfo  {journal}
  {Phys. Rev. Lett.} }\textbf {\bibinfo {volume} {110}},\ \bibinfo {pages}
  {066802} (\bibinfo {year} {2013})}\BibitemShut {NoStop}%
\bibitem [{\citenamefont {Wallraff}\ \emph {et~al.}(2013)\citenamefont
  {Wallraff}, \citenamefont {Stockklauser}, \citenamefont {Ihn}, \citenamefont
  {Petta},\ and\ \citenamefont {Blais}}]{Wallraff2013}%
  \BibitemOpen
  \bibfield  {author} {\bibinfo {author} {\bibfnamefont {A.}~\bibnamefont
  {Wallraff}}, \bibinfo {author} {\bibfnamefont {A.}~\bibnamefont
  {Stockklauser}}, \bibinfo {author} {\bibfnamefont {T.}~\bibnamefont {Ihn}},
  \bibinfo {author} {\bibfnamefont {J.~R.}\ \bibnamefont {Petta}}, \ and\
  \bibinfo {author} {\bibfnamefont {A.}~\bibnamefont {Blais}},\ }\Doi
  {10.1103/PhysRevLett.111.249701} {\bibfield  {journal} {\bibinfo  {journal}
  {Phys. Rev. Lett.} }\textbf {\bibinfo {volume} {111}},\ \bibinfo {pages}
  {249701} (\bibinfo {year} {2013})}\BibitemShut {NoStop}%
\bibitem [{\citenamefont {Basset}\ \emph {et~al.}(2013)\citenamefont {Basset},
  \citenamefont {Jarausch}, \citenamefont {Stockklauser}, \citenamefont {Frey},
  \citenamefont {Reichl}, \citenamefont {Wegscheider}, \citenamefont {Ihn},
  \citenamefont {Ensslin},\ and\ \citenamefont {Wallraff}}]{Basset2013}%
  \BibitemOpen
  \bibfield  {author} {\bibinfo {author} {\bibfnamefont {J.}~\bibnamefont
  {Basset}}, \bibinfo {author} {\bibfnamefont {D.-D.}\ \bibnamefont
  {Jarausch}}, \bibinfo {author} {\bibfnamefont {A.}~\bibnamefont
  {Stockklauser}}, \bibinfo {author} {\bibfnamefont {T.}~\bibnamefont {Frey}},
  \bibinfo {author} {\bibfnamefont {C.}~\bibnamefont {Reichl}}, \bibinfo
  {author} {\bibfnamefont {W.}~\bibnamefont {Wegscheider}}, \bibinfo {author}
  {\bibfnamefont {T.}~\bibnamefont {Ihn}}, \bibinfo {author} {\bibfnamefont
  {K.}~\bibnamefont {Ensslin}}, \ and\ \bibinfo {author} {\bibfnamefont
  {A.}~\bibnamefont {Wallraff}},\ }\Doi
  {10.1103/PhysRevB.88.125312} {\bibfield  {journal} {\bibinfo  {journal}
  {Phys. Rev. B} }\textbf {\bibinfo {volume} {88}},\ \bibinfo {pages}
  {125312} (\bibinfo {year} {2013})}\BibitemShut {NoStop}%
\bibitem [{\citenamefont {Trif}\ \emph {et~al.}(2008)\citenamefont {Trif},
  \citenamefont {Golovach},\ and\ \citenamefont {Loss}}]{Trif2008}%
  \BibitemOpen
  \bibfield  {author} {\bibinfo {author} {\bibfnamefont {M.}~\bibnamefont
  {Trif}}, \bibinfo {author} {\bibfnamefont {V.~N.}\ \bibnamefont {Golovach}},
  \ and\ \bibinfo {author} {\bibfnamefont {D.}~\bibnamefont {Loss}},\ }\Doi
  {10.1103/PhysRevB.77.045434} {\bibfield  {journal} {\bibinfo  {journal}
  {Phys. Rev. B} }\textbf {\bibinfo {volume} {77}},\ \bibinfo {pages}
  {045434} (\bibinfo {year} {2008})}\BibitemShut {NoStop}%
\bibitem [{\citenamefont {Cottet}\ \emph {et~al.}(2012)\citenamefont {Cottet},
  \citenamefont {Kontos},\ and\ \citenamefont {Yeyati}}]{Cottet2012}%
  \BibitemOpen
  \bibfield  {author} {\bibinfo {author} {\bibfnamefont {A.}~\bibnamefont
  {Cottet}}, \bibinfo {author} {\bibfnamefont {T.}~\bibnamefont {Kontos}}, \
  and\ \bibinfo {author} {\bibfnamefont {A.~L.}\ \bibnamefont {Yeyati}},\ }\Doi
  {10.1103/PhysRevLett.108.166803} {\bibfield  {journal} {\bibinfo  {journal}
  {Phys. Rev. Lett.} }\textbf {\bibinfo {volume} {108}},\ \bibinfo {pages}
  {166803} (\bibinfo {year} {2012})}\BibitemShut {NoStop}%
\bibitem [{\citenamefont {Childress}\ \emph {et~al.}(2004)\citenamefont
  {Childress}, \citenamefont {S\o{}rensen},\ and\ \citenamefont
  {Lukin}}]{Childress2004}%
  \BibitemOpen
  \bibfield  {author} {\bibinfo {author} {\bibfnamefont {L.}~\bibnamefont
  {Childress}}, \bibinfo {author} {\bibfnamefont {A.~S.}\ \bibnamefont
  {S\o{}rensen}}, \ and\ \bibinfo {author} {\bibfnamefont {M.~D.}\ \bibnamefont
  {Lukin}},\ }\Doi {10.1103/PhysRevA.69.042302} {\bibfield  {journal} {\bibinfo
   {journal} {Phys. Rev. A} }\textbf {\bibinfo {volume} {69}},\ \bibinfo
  {pages} {042302} (\bibinfo {year} {2004})}\BibitemShut {NoStop}%
\bibitem [{\citenamefont {Jin}\ \emph {et~al.}(2011)\citenamefont {Jin},
  \citenamefont {Marthaler}, \citenamefont {Cole}, \citenamefont {Shnirman},\
  and\ \citenamefont {Sch\"on}}]{Jin2011}%
  \BibitemOpen
  \bibfield  {author} {\bibinfo {author} {\bibfnamefont {P.-Q.}\ \bibnamefont
  {Jin}}, \bibinfo {author} {\bibfnamefont {M.}~\bibnamefont {Marthaler}},
  \bibinfo {author} {\bibfnamefont {J.~H.}\ \bibnamefont {Cole}}, \bibinfo
  {author} {\bibfnamefont {A.}~\bibnamefont {Shnirman}}, \ and\ \bibinfo
  {author} {\bibfnamefont {G.}~\bibnamefont {Sch\"on}},\ }\Doi
  {10.1103/PhysRevB.84.035322} {\bibfield  {journal} {\bibinfo  {journal}
  {Phys. Rev. B} }\textbf {\bibinfo {volume} {84}},\ \bibinfo {pages}
  {035322} (\bibinfo {year} {2011})}\BibitemShut {NoStop}%
\bibitem [{\citenamefont {Bergenfeldt}\ and\ \citenamefont
  {Samuelsson}(2012)}]{Bergenfeldt2012}%
  \BibitemOpen
  \bibfield  {author} {\bibinfo {author} {\bibfnamefont {C.}~\bibnamefont
  {Bergenfeldt}}\ and\ \bibinfo {author} {\bibfnamefont {P.}~\bibnamefont
  {Samuelsson}},\ }\Doi {10.1103/PhysRevB.85.045446} {\bibfield  {journal}
  {\bibinfo  {journal} {Phys. Rev. B} }\textbf {\bibinfo {volume} {85}},\
  \bibinfo {pages} {045446} (\bibinfo {year} {2012})}\BibitemShut {NoStop}%
\bibitem [{\citenamefont {Bergenfeldt}\ and\ \citenamefont
  {Samuelsson}(2013)}]{Bergenfeldt2013}%
  \BibitemOpen
  \bibfield  {author} {\bibinfo {author} {\bibfnamefont {C.}~\bibnamefont
  {Bergenfeldt}}\ and\ \bibinfo {author} {\bibfnamefont {P.}~\bibnamefont
  {Samuelsson}},\ }\Doi {10.1103/PhysRevB.87.195427} {\bibfield  {journal}
  {\bibinfo  {journal} {Phys. Rev. B} }\textbf {\bibinfo {volume} {87}},\
  \bibinfo {pages} {195427} (\bibinfo {year} {2013})}\BibitemShut {NoStop}%
\bibitem [{\citenamefont {Lambert}\ \emph {et~al.}(2013)\citenamefont
  {Lambert}, \citenamefont {Flindt},\ and\ \citenamefont {Nori}}]{Lambert2013}%
  \BibitemOpen
  \bibfield  {author} {\bibinfo {author} {\bibfnamefont {N.}~\bibnamefont
  {Lambert}}, \bibinfo {author} {\bibfnamefont {C.}~\bibnamefont {Flindt}}, \
  and\ \bibinfo {author} {\bibfnamefont {F.}~\bibnamefont {Nori}},\ }\Doi {10.1209/0295-5075/103/17005}
  {\bibfield  {journal} {\bibinfo  {journal} {Europhys. Lett.} }\textbf
  {\bibinfo {volume} {103}},\ \bibinfo {pages} {17005} (\bibinfo {year}
  {2013})}\BibitemShut {NoStop}%
\bibitem [{\citenamefont {Contreras-Pulido}\ \emph {et~al.}(2013)\citenamefont
  {Contreras-Pulido}, \citenamefont {Emary}, \citenamefont {Brandes},\ and\
  \citenamefont {Aguado}}]{Pulido2013}%
  \BibitemOpen
  \bibfield  {author} {\bibinfo {author} {\bibfnamefont {L.~D.}\ \bibnamefont
  {Contreras-Pulido}}, \bibinfo {author} {\bibfnamefont {C.}~\bibnamefont
  {Emary}}, \bibinfo {author} {\bibfnamefont {T.}~\bibnamefont {Brandes}}, \
  and\ \bibinfo {author} {\bibfnamefont {R.}~\bibnamefont {Aguado}},\
  }\Doi {10.1088/1367-2630/15/9/095008} {\bibfield  {journal}
  {\bibinfo  {journal} {New J. Phys.} }\textbf {\bibinfo {volume} {15}},\
  \bibinfo {pages} {095008} (\bibinfo {year} {2013})}\BibitemShut {NoStop}%
\bibitem [{\citenamefont {Xu}\ and\ \citenamefont
  {Vavilov}(2013){\natexlab{a}}}]{Xu2013}%
  \BibitemOpen
  \bibfield  {author} {\bibinfo {author} {\bibfnamefont {C.}~\bibnamefont
  {Xu}}\ and\ \bibinfo {author} {\bibfnamefont {M.~G.}\ \bibnamefont
  {Vavilov}},\ }\Doi {10.1103/PhysRevB.87.035429} {\bibfield  {journal}
  {\bibinfo  {journal} {Phys. Rev. B} }\textbf {\bibinfo {volume} {87}},\
  \bibinfo {pages} {035429} (\bibinfo {year} {2013}{\natexlab{a}})}\BibitemShut
  {NoStop}%
\bibitem [{\citenamefont {Xu}\ and\ \citenamefont
  {Vavilov}(2013){\natexlab{b}}}]{Xu2013b}%
  \BibitemOpen
  \bibfield  {author} {\bibinfo {author} {\bibfnamefont {C.}~\bibnamefont
  {Xu}}\ and\ \bibinfo {author} {\bibfnamefont {M.~G.}\ \bibnamefont
  {Vavilov}},\ }\Doi {10.1103/PhysRevB.88.195307} {\bibfield  {journal}
  {\bibinfo  {journal} {Phys. Rev. B} }\textbf {\bibinfo {volume} {88}},\
  \bibinfo {pages} {195307} (\bibinfo {year} {2013})}\BibitemShut {NoStop}%
\bibitem [{\citenamefont {Beenakker}\ and\ \citenamefont
  {Staring}(1992)}]{Beenakker1992}%
  \BibitemOpen
  \bibfield  {author} {\bibinfo {author} {\bibfnamefont {C.~W.~J.}\
  \bibnamefont {Beenakker}}\ and\ \bibinfo {author} {\bibfnamefont {A.~A.~M.}\
  \bibnamefont {Staring}},\ }\Doi {10.1103/PhysRevB.46.9667} {\bibfield
  {journal} {\bibinfo  {journal} {Phys. Rev. B} }\textbf {\bibinfo {volume}
  {46}},\ \bibinfo {pages} {9667} (\bibinfo {year} {1992})}\BibitemShut
  {NoStop}%
\bibitem [{\citenamefont {Humphrey}\ \emph {et~al.}(2002)\citenamefont
  {Humphrey}, \citenamefont {Newbury}, \citenamefont {Taylor},\ and\
  \citenamefont {Linke}}]{Humphrey2002}%
  \BibitemOpen
  \bibfield  {author} {\bibinfo {author} {\bibfnamefont {T.~E.}\ \bibnamefont
  {Humphrey}}, \bibinfo {author} {\bibfnamefont {R.}~\bibnamefont {Newbury}},
  \bibinfo {author} {\bibfnamefont {R.~P.}\ \bibnamefont {Taylor}}, \ and\
  \bibinfo {author} {\bibfnamefont {H.}~\bibnamefont {Linke}},\ }\Doi
  {10.1103/PhysRevLett.89.116801} {\bibfield  {journal} {\bibinfo  {journal}
  {Phys. Rev. Lett.} }\textbf {\bibinfo {volume} {89}},\ \bibinfo {pages}
  {116801} (\bibinfo {year} {2002})}\BibitemShut {NoStop}%
\bibitem [{\citenamefont {Esposito}\ \emph {et~al.}(2009)\citenamefont
  {Esposito}, \citenamefont {Lindenberg},\ and\ \citenamefont {Van~den
  Broeck}}]{Esposito2009}%
  \BibitemOpen
  \bibfield  {author} {\bibinfo {author} {\bibfnamefont {M.}~\bibnamefont
  {Esposito}}, \bibinfo {author} {\bibfnamefont {K.}~\bibnamefont
  {Lindenberg}}, \ and\ \bibinfo {author} {\bibfnamefont {C.}~\bibnamefont
  {Van~den Broeck}},\ }\Doi {10.1209/0295-5075/85/60010} {\bibfield  {journal}
  {\bibinfo  {journal} {Europhys. Lett.} }\textbf {\bibinfo {volume} {85}},\
  \bibinfo {pages} {60010} (\bibinfo {year} {2009})}\BibitemShut {NoStop}%
\bibitem [{\citenamefont {Nakpathomkun}\ \emph {et~al.}(2010)\citenamefont
  {Nakpathomkun}, \citenamefont {Xu},\ and\ \citenamefont
  {Linke}}]{Nakpathomkun2010}%
  \BibitemOpen
  \bibfield  {author} {\bibinfo {author} {\bibfnamefont {N.}~\bibnamefont
  {Nakpathomkun}}, \bibinfo {author} {\bibfnamefont {H.~Q.}\ \bibnamefont
  {Xu}}, \ and\ \bibinfo {author} {\bibfnamefont {H.}~\bibnamefont {Linke}},\
  }\Doi {10.1103/PhysRevB.82.235428} {\bibfield  {journal} {\bibinfo  {journal}
  {Phys. Rev. B} }\textbf {\bibinfo {volume} {82}},\ \bibinfo {pages}
  {235428} (\bibinfo {year} {2010})}\BibitemShut {NoStop}%
\bibitem [{\citenamefont {Entin-Wohlman}\ \emph {et~al.}(2010)\citenamefont
  {Entin-Wohlman}, \citenamefont {Imry},\ and\ \citenamefont
  {Aharony}}]{Entin2010}%
  \BibitemOpen
  \bibfield  {author} {\bibinfo {author} {\bibfnamefont {O.}~\bibnamefont
  {Entin-Wohlman}}, \bibinfo {author} {\bibfnamefont {Y.}~\bibnamefont {Imry}},
  \ and\ \bibinfo {author} {\bibfnamefont {A.}~\bibnamefont {Aharony}},\ }\Doi
  {10.1103/PhysRevB.82.115314} {\bibfield  {journal} {\bibinfo  {journal}
  {Phys. Rev. B} }\textbf {\bibinfo {volume} {82}},\ \bibinfo {pages}
  {115314} (\bibinfo {year} {2010})}\BibitemShut {NoStop}%
\bibitem [{\citenamefont {S\'anchez}\ and\ \citenamefont
  {B\"{u}ttiker}(2011)}]{Sanchez2011}%
  \BibitemOpen
  \bibfield  {author} {\bibinfo {author} {\bibfnamefont {R.}~\bibnamefont
  {S\'anchez}}\ and\ \bibinfo {author} {\bibfnamefont {M.}~\bibnamefont
  {B\"{u}ttiker}},\ }\Doi {10.1103/PhysRevB.83.085428} {\bibfield  {journal}
  {\bibinfo  {journal} {Phys. Rev. B} }\textbf {\bibinfo {volume} {83}},\
  \bibinfo {pages} {085428} (\bibinfo {year} {2011})}\BibitemShut {NoStop}%
\bibitem [{\citenamefont {Sothmann}\ and\ \citenamefont
  {B\"{u}ttiker}(2012)}]{Sothmann2012}%
  \BibitemOpen
  \bibfield  {author} {\bibinfo {author} {\bibfnamefont {B.}~\bibnamefont
  {Sothmann}}\ and\ \bibinfo {author} {\bibfnamefont {M.}~\bibnamefont
  {B\"{u}ttiker}},\ }\Doi {10.1209/0295-5075/99/27001} {\bibfield  {journal}
  {\bibinfo  {journal} {Europhys. Lett.} }\textbf {\bibinfo {volume} {99}},\
  \bibinfo {pages} {27001} (\bibinfo {year} {2012})}\BibitemShut {NoStop}%
\bibitem [{\citenamefont {Sothmann}\ \emph {et~al.}(2012)\citenamefont
  {Sothmann}, \citenamefont {S\'anchez}, \citenamefont {Jordan},\ and\
  \citenamefont {B\"uttiker}}]{Sothmann2012b}%
  \BibitemOpen
  \bibfield  {author} {\bibinfo {author} {\bibfnamefont {B.}~\bibnamefont
  {Sothmann}}, \bibinfo {author} {\bibfnamefont {R.}~\bibnamefont {S\'anchez}},
  \bibinfo {author} {\bibfnamefont {A.~N.}\ \bibnamefont {Jordan}}, \ and\
  \bibinfo {author} {\bibfnamefont {M.}~\bibnamefont {B\"uttiker}},\ }\Doi
  {10.1103/PhysRevB.85.205301} {\bibfield  {journal} {\bibinfo  {journal}
  {Phys. Rev. B} }\textbf {\bibinfo {volume} {85}},\ \bibinfo {pages}
  {205301} (\bibinfo {year} {2012})}\BibitemShut {NoStop}%
\bibitem [{\citenamefont {Ruokola}\ and\ \citenamefont
  {Ojanen}(2012)}]{Ruokola2012}%
  \BibitemOpen
  \bibfield  {author} {\bibinfo {author} {\bibfnamefont {T.}~\bibnamefont
  {Ruokola}}\ and\ \bibinfo {author} {\bibfnamefont {T.}~\bibnamefont
  {Ojanen}},\ }\Doi {10.1103/PhysRevB.86.035454} {\bibfield  {journal}
  {\bibinfo  {journal} {Phys. Rev. B} }\textbf {\bibinfo {volume} {86}},\
  \bibinfo {pages} {035454} (\bibinfo {year} {2012})}\BibitemShut {NoStop}%
\bibitem [{\citenamefont {Jordan}\ \emph {et~al.}(2013)\citenamefont {Jordan},
  \citenamefont {Sothmann}, \citenamefont {S\'anchez},\ and\ \citenamefont
  {B\"{u}ttiker}}]{Jordan2013}%
  \BibitemOpen
  \bibfield  {author} {\bibinfo {author} {\bibfnamefont {A.~N.}\ \bibnamefont
  {Jordan}}, \bibinfo {author} {\bibfnamefont {B.}~\bibnamefont {Sothmann}},
  \bibinfo {author} {\bibfnamefont {R.}~\bibnamefont {S\'anchez}}, \ and\
  \bibinfo {author} {\bibfnamefont {M.}~\bibnamefont {B\"{u}ttiker}},\ }\Doi
  {10.1103/PhysRevB.87.075312} {\bibfield  {journal} {\bibinfo  {journal}
  {Phys. Rev. B} }\textbf {\bibinfo {volume} {87}},\ \bibinfo {pages}
  {075312} (\bibinfo {year} {2013})}\BibitemShut {NoStop}%
\bibitem [{\citenamefont {Kennes}\ \emph {et~al.}(2013)\citenamefont {Kennes},
  \citenamefont {Schuricht},\ and\ \citenamefont {Meden}}]{Kennes2013}%
  \BibitemOpen
  \bibfield  {author} {\bibinfo {author} {\bibfnamefont {D.~M.}\ \bibnamefont
  {Kennes}}, \bibinfo {author} {\bibfnamefont {D.}~\bibnamefont {Schuricht}}, \
  and\ \bibinfo {author} {\bibfnamefont {V.}~\bibnamefont {Meden}},\ }\Doi
  {10.1209/0295-5075/102/57003} {\bibfield  {journal} {\bibinfo  {journal}
  {Europhys. Lett.} }\textbf {\bibinfo {volume} {102}},\ \bibinfo {pages}
  {57003} (\bibinfo {year} {2013})}\BibitemShut {NoStop}%
\bibitem [{\citenamefont {Staring}\ \emph {et~al.}(1993)\citenamefont
  {Staring}, \citenamefont {Molenkamp}, \citenamefont {Alphenaar},
  \citenamefont {van Houten}, \citenamefont {Buyk}, \citenamefont {Mabesoone},
  \citenamefont {Beenakker},\ and\ \citenamefont {Foxon}}]{Staring1993}%
  \BibitemOpen
  \bibfield  {author} {\bibinfo {author} {\bibfnamefont {A.~A.~M.}\
  \bibnamefont {Staring}}, \bibinfo {author} {\bibfnamefont {L.~W.}\
  \bibnamefont {Molenkamp}}, \bibinfo {author} {\bibfnamefont {B.~W.}\
  \bibnamefont {Alphenaar}}, \bibinfo {author} {\bibfnamefont {H.}~\bibnamefont
  {van Houten}}, \bibinfo {author} {\bibfnamefont {O.~J.~A.}\ \bibnamefont
  {Buyk}}, \bibinfo {author} {\bibfnamefont {M.~A.~A.}\ \bibnamefont
  {Mabesoone}}, \bibinfo {author} {\bibfnamefont {C.~W.~J.}\ \bibnamefont
  {Beenakker}}, \ and\ \bibinfo {author} {\bibfnamefont {C.~T.}\ \bibnamefont
  {Foxon}},\ }\Doi {10.1209/0295-5075/22/1/011} {\bibfield  {journal} {\bibinfo
   {journal} {Europhys. Lett.} }\textbf {\bibinfo {volume} {22}},\ \bibinfo
  {pages} {57} (\bibinfo {year} {1993})}\BibitemShut {NoStop}%
\bibitem [{\citenamefont {Dzurak}\ \emph {et~al.}(1997)\citenamefont {Dzurak},
  \citenamefont {Smith}, \citenamefont {Barnes}, \citenamefont {Pepper},
  \citenamefont {Mart\'{\i}n-Moreno}, \citenamefont {Liang}, \citenamefont
  {Ritchie},\ and\ \citenamefont {Jones}}]{Dzurak1997}%
  \BibitemOpen
  \bibfield  {author} {\bibinfo {author} {\bibfnamefont {A.~S.}\ \bibnamefont
  {Dzurak}}, \bibinfo {author} {\bibfnamefont {C.~G.}\ \bibnamefont {Smith}},
  \bibinfo {author} {\bibfnamefont {C.~H.~W.}\ \bibnamefont {Barnes}}, \bibinfo
  {author} {\bibfnamefont {M.}~\bibnamefont {Pepper}}, \bibinfo {author}
  {\bibfnamefont {L.}~\bibnamefont {Martin-Moreno}}, \bibinfo {author}
  {\bibfnamefont {C.~T.}\ \bibnamefont {Liang}}, \bibinfo {author}
  {\bibfnamefont {D.~A.}\ \bibnamefont {Ritchie}}, \ and\ \bibinfo {author}
  {\bibfnamefont {G.~A.~C.}\ \bibnamefont {Jones}},\ }\Doi
  {10.1103/PhysRevB.55.R10197} {\bibfield  {journal} {\bibinfo  {journal}
  {Phys. Rev. B} }\textbf {\bibinfo {volume} {55}},\ \bibinfo {pages}
  {R10197} (\bibinfo {year} {1997})}\BibitemShut {NoStop}%
\bibitem [{\citenamefont {Scheibner}\ \emph {et~al.}(2005)\citenamefont
  {Scheibner}, \citenamefont {Buhmann}, \citenamefont {Reuter}, \citenamefont
  {Kiselev},\ and\ \citenamefont {Molenkamp}}]{Scheibner2005}%
  \BibitemOpen
  \bibfield  {author} {\bibinfo {author} {\bibfnamefont {R.}~\bibnamefont
  {Scheibner}}, \bibinfo {author} {\bibfnamefont {H.}~\bibnamefont {Buhmann}},
  \bibinfo {author} {\bibfnamefont {D.}~\bibnamefont {Reuter}}, \bibinfo
  {author} {\bibfnamefont {M.~N.}\ \bibnamefont {Kiselev}}, \ and\ \bibinfo
  {author} {\bibfnamefont {L.~W.}\ \bibnamefont {Molenkamp}},\ }\Doi
  {10.1103/PhysRevLett.95.176602} {\bibfield  {journal} {\bibinfo  {journal}
  {Phys. Rev. Lett.} }\textbf {\bibinfo {volume} {95}},\ \bibinfo {pages}
  {176602} (\bibinfo {year} {2005})}\BibitemShut {NoStop}%
\bibitem [{\citenamefont {Scheibner}\ \emph {et~al.}(2007)\citenamefont
  {Scheibner}, \citenamefont {Novik}, \citenamefont {Borzenko}, \citenamefont
  {König}, \citenamefont {Reuter}, \citenamefont {Wieck}, \citenamefont
  {Buhmann},\ and\ \citenamefont {Molenkamp}}]{Scheibner2007}%
  \BibitemOpen
  \bibfield  {author} {\bibinfo {author} {\bibfnamefont {R.}~\bibnamefont
  {Scheibner}}, \bibinfo {author} {\bibfnamefont {E.~G.}\ \bibnamefont
  {Novik}}, \bibinfo {author} {\bibfnamefont {T.}~\bibnamefont {Borzenko}},
  \bibinfo {author} {\bibfnamefont {M.}~\bibnamefont {Konig}}, \bibinfo
  {author} {\bibfnamefont {D.}~\bibnamefont {Reuter}}, \bibinfo {author}
  {\bibfnamefont {A.~D.}\ \bibnamefont {Wieck}}, \bibinfo {author}
  {\bibfnamefont {H.}~\bibnamefont {Buhmann}}, \ and\ \bibinfo {author}
  {\bibfnamefont {L.~W.}\ \bibnamefont {Molenkamp}},\ }\Doi
  {10.1103/PhysRevB.75.041301} {\bibfield  {journal} {\bibinfo  {journal}
  {Phys. Rev. B} }\textbf {\bibinfo {volume} {75}},\ \bibinfo {pages}
  {041301} (\bibinfo {year} {2007})}\BibitemShut {NoStop}%
\bibitem [{\citenamefont {Svensson}\ \emph {et~al.}(2012)\citenamefont
  {Svensson}, \citenamefont {Persson}, \citenamefont {Hoffmann}, \citenamefont
  {Nakpathomkun}, \citenamefont {Nilsson}, \citenamefont {Xu}, \citenamefont
  {Samuelson},\ and\ \citenamefont {Linke}}]{Svensson2012}%
  \BibitemOpen
  \bibfield  {author} {\bibinfo {author} {\bibfnamefont {S.~F.}\ \bibnamefont
  {Svensson}}, \bibinfo {author} {\bibfnamefont {A.~I.}\ \bibnamefont
  {Persson}}, \bibinfo {author} {\bibfnamefont {E.~A.}\ \bibnamefont
  {Hoffmann}}, \bibinfo {author} {\bibfnamefont {N.}~\bibnamefont
  {Nakpathomkun}}, \bibinfo {author} {\bibfnamefont {H.~A.}\ \bibnamefont
  {Nilsson}}, \bibinfo {author} {\bibfnamefont {H.~Q.}\ \bibnamefont {Xu}},
  \bibinfo {author} {\bibfnamefont {L.}~\bibnamefont {Samuelson}}, \ and\
  \bibinfo {author} {\bibfnamefont {H.}~\bibnamefont {Linke}},\ }\Doi
  {10.1088/1367-2630/14/3/033041} {\bibfield  {journal} {\bibinfo  {journal}
  {New J. Phys.} }\textbf {\bibinfo {volume} {14}},\ \bibinfo {pages}
  {033041} (\bibinfo {year} {2012})}\BibitemShut {NoStop}%
\bibitem [{\citenamefont {Dresselhaus}\ \emph {et~al.}(2007)\citenamefont {Dresselhaus},
  \citenamefont {Chen}, \citenamefont {Tang},\citenamefont {Yang}, \citenamefont {Lee}, \citenamefont {Wang}, \citenamefont {Ren}, \citenamefont {Fleurial}\ and\ \citenamefont
  {Gogna}}]{Dresselhaus2007}%
  \BibitemOpen
  \bibfield  {author} {\bibinfo {author} {\bibfnamefont {M.~S.}\ \bibnamefont
  {Dresselhaus}}, \bibinfo {author} {\bibfnamefont {G.}~\bibnamefont {Chen}},
  \bibinfo {author} {\bibfnamefont {M.~Y.}~\bibnamefont {Tang}},
   \bibinfo {author} {\bibfnamefont {R.~G.}~\bibnamefont {Yang}},
  \bibinfo {author} {\bibfnamefont {H.}~\bibnamefont {Lee}},
  \bibinfo {author} {\bibfnamefont {D.~Z.}~\bibnamefont {Wang}},
  \bibinfo {author} {\bibfnamefont {Z.~F.}~\bibnamefont {Ren}},
    \bibinfo {author} {\bibfnamefont {J.-P.}~\bibnamefont {Fleurial}}
  \ and\
  \bibinfo {author} {\bibfnamefont {P.}~\bibnamefont {Gogna}},\ }\Doi
  {10.1002/adma.200600527} {\bibfield  {journal} {\bibinfo  {journal}
  {Adv. Mater.} }\textbf {\bibinfo {volume} {19}},\ \bibinfo {pages}
  {1043} (\bibinfo {year} {2007})}\BibitemShut {NoStop}%
\bibitem [{\citenamefont {Meschke}\ \emph {et~al.}(2006)\citenamefont
  {Meschke}, \citenamefont {Guichard},\ and\ \citenamefont
  {Pekola}}]{Meschke2006}%
  \BibitemOpen
  \bibfield  {author} {\bibinfo {author} {\bibfnamefont {M.}~\bibnamefont
  {Meschke}}, \bibinfo {author} {\bibfnamefont {W.}~\bibnamefont {Guichard}}, \
  and\ \bibinfo {author} {\bibfnamefont {J.~P.}\ \bibnamefont {Pekola}},\ }\Doi
  {10.1038/nature05276} {\bibfield  {journal} {\bibinfo  {journal} {Nature}
  }\textbf {\bibinfo {volume} {444}},\ \bibinfo {pages} {187} (\bibinfo {year}
  {2006})}\BibitemShut {NoStop}%
\bibitem [{\citenamefont {Savin}\ \emph {et~al.}(2006)\citenamefont {Savin},
  \citenamefont {Pekola}, \citenamefont {Averin},\ and\ \citenamefont
  {Semenov}}]{Savin2006}%
  \BibitemOpen
  \bibfield  {author} {\bibinfo {author} {\bibfnamefont {A.~M.}\ \bibnamefont
  {Savin}}, \bibinfo {author} {\bibfnamefont {J.~P.}~\bibnamefont {Pekola}},
  \bibinfo {author} {\bibfnamefont {D.~V.}~\bibnamefont {Averin}}, \ and\
  \bibinfo {author} {\bibfnamefont {V.~K.}~\bibnamefont {Semenow}},\ }\Doi
  {10.1063/1.2187276} {\bibfield  {journal} {\bibinfo  {journal}
  {J. Appl. Phys.} }\textbf {\bibinfo {volume} {99}},\ \bibinfo {pages}
  {084501} (\bibinfo {year} {2006})}\BibitemShut {NoStop}%
\bibitem [{\citenamefont {Breuer}\ and\ \citenamefont
  {Petruccione}(2002)}]{Breuer2002}%
  \BibitemOpen
  \bibfield  {author} {\bibinfo {author} {\bibfnamefont {H.-P.}\ \bibnamefont
  {Breuer}}\ and\ \bibinfo {author} {\bibfnamefont {F.}~\bibnamefont
  {Petruccione}},\ }\href@noop {} {\emph {\bibinfo {title} {The Theory of Open
  Quantum Systems}}}\ (\bibinfo  {publisher} {Oxford University Press},\
  \bibinfo {year} {2002})\BibitemShut {NoStop}%
\bibitem [{\citenamefont {Wang}\ \emph {et~al.}(2009)\citenamefont {Wang},
  \citenamefont {Hofheinz}, \citenamefont {Wenner},\ \citenamefont {Ansmann},\ \citenamefont {Bialczak},\
  \citenamefont {Lenander},\ \citenamefont {Lucero},\ \citenamefont {Neeley},\ \citenamefont {Lucero},\ \citenamefont {O'Connell},\
    \citenamefont {Sank},\ \citenamefont {Weides},\ \citenamefont {Sank},\ \citenamefont {Cleland},\ and\ \citenamefont{Martinis}}]{Wang2009}%
  \BibitemOpen
  \bibfield  {author} {\bibinfo {author} {\bibfnamefont {H.}\ \bibnamefont {Wang}}, \bibinfo {author} {\bibfnamefont {M.}~\bibnamefont {Hofheinz}},
  \bibinfo {author} {\bibfnamefont {J.}~\bibnamefont {Wenner}},  \bibinfo {author} {\bibfnamefont {M.}~\bibnamefont {Ansmann}},
  \bibinfo {author} {\bibfnamefont {R.~C.}~\bibnamefont {Bialczak}},  \bibinfo {author} {\bibfnamefont {M.}~\bibnamefont {Lenander}},
   \bibinfo {author} {\bibfnamefont {E.}~\bibnamefont {Lucero}}, \bibinfo {author} {\bibfnamefont {M.}~\bibnamefont {Neeley}},
   \bibinfo {author} {\bibfnamefont {A.~D.}~\bibnamefont {O'Connell}}, \bibinfo {author} {\bibfnamefont {D.}~\bibnamefont {Sank}}
    \bibinfo {author} {\bibfnamefont {M.}~\bibnamefont {Weides}}, \bibinfo {author} {\bibfnamefont {A.~N.}~\bibnamefont {Cleland}},\ and\ \bibinfo {author} {\bibfnamefont {J.~M.}~\bibnamefont {Martinis}},\ }\Doi
  {10.1063/1.3273372} {\bibfield  {journal} {\bibinfo  {journal}
  {Appl. Phys. Lett.} }\textbf {\bibinfo {volume} {95}},\ \bibinfo {pages}
  {233508} (\bibinfo {year} {2009})}\BibitemShut {NoStop}%
\bibitem [{Note1()}]{Note1}%
  \BibitemOpen
  \bibinfo {note} {The case $\theta _2\rightarrow \pi /2$ should be treated
  with care as the effective couplings to the cavity vanish for $\theta _2=\pi
  /2$ and we go outside the strong coupling regime assumed so far.}\BibitemShut
  {Stop}%
\bibitem [{\citenamefont {Van~den Broeck}(2005)}]{Broeck2005}%
  \BibitemOpen
  \bibfield  {author} {\bibinfo {author} {\bibfnamefont {C.}~\bibnamefont
  {Van~den Broeck}},\ }\Doi {10.1103/PhysRevLett.95.190602} {\bibfield
  {journal} {\bibinfo  {journal} {Phys. Rev. Lett.} }\textbf {\bibinfo
  {volume} {95}},\ \bibinfo {pages} {190602} (\bibinfo {year}
  {2005})}\BibitemShut {NoStop}%
\bibitem [{\citenamefont {Schmiedl}\ and\ \citenamefont
  {Seifert}(2008)}]{Schmiedl2008}%
  \BibitemOpen
  \bibfield  {author} {\bibinfo {author} {\bibfnamefont {T.}~\bibnamefont
  {Schmiedl}}\ and\ \bibinfo {author} {\bibfnamefont {U.}~\bibnamefont
  {Seifert}},\ }\Doi {10.1209/0295-5075/81/20003} {\bibfield  {journal}
  {\bibinfo  {journal} {Europhys. Lett.} }\textbf {\bibinfo {volume} {81}},\
  \bibinfo {pages} {20003} (\bibinfo {year} {2008})}\BibitemShut {NoStop}%
\bibitem [{\citenamefont {Hayashi}\ \emph {et~al.}(2003)\citenamefont
  {Hayashi}, \citenamefont {Fujisawa}, \citenamefont {Cheong}, \citenamefont
  {Jeong},\ and\ \citenamefont {Hirayama}}]{Hayashi2003}%
  \BibitemOpen
  \bibfield  {author} {\bibinfo {author} {\bibfnamefont {T.}~\bibnamefont
  {Hayashi}}, \bibinfo {author} {\bibfnamefont {T.}~\bibnamefont {Fujisawa}},
  \bibinfo {author} {\bibfnamefont {H.~D.}~\bibnamefont {Cheong}}, \bibinfo
  {author} {\bibfnamefont {Y.~H.}\ \bibnamefont {Jeong}}, \ and\ \bibinfo
  {author} {\bibfnamefont {Y.}\ \bibnamefont {Hirayama}},\ }\Doi
  {10.1103/PhysRevLett.91.226804} {\bibfield  {journal} {\bibinfo  {journal}
  {Phys. Rev. Lett.} }\textbf {\bibinfo {volume} {91}},\ \bibinfo {pages}
  {226804} (\bibinfo {year} {2003})}\BibitemShut {NoStop}%
\end{thebibliography}
\end{document}